\newcommand{\stdpack}{
  \usepackage{amssymb}
  \usepackage{amsmath}
  \usepackage{eucal}
  \usepackage[final]{graphicx}
  \usepackage{psfrag}
  \usepackage{fancyhdr}
  \renewcommand{\headrulewidth}{0pt}\lhead{}\cfoot{}\rfoot{\thepage}

  \newcommand{\draft}{\usepackage[light,first]{draftcopy}\draftcopyName{draft}{350}}
  \newcommand{\labels}{\usepackage{showlabels}}
  \newcommand{\maple}{\usepackage{maple2e}}
  \newcommand{\makeidx}{\usepackage{makeidx}\makeindex}
}
\newcommand{\std}[1]{
  \stdpack
  \usepackage{a4,a4wide}
  \newcommand{\blockindent}{3ex}
  \renewcommand{\baselinestretch}{#1}
  \renewcommand{\arraystretch}{1.2}
  \hoffset -1cm  \addtolength{\textwidth}{2cm}
  \voffset -0cm  \addtolength{\textheight}{0cm}

  \usepackage{./mt}
  \columnsep 5ex
  \parindent 3ex
  \parskip 1ex
  \macros
  \pagestyle{fancy}
  \bibliographystyle{./chicago}
\renewenvironment{abstract}{\paragraph{Abstract}\begin{rblock}\small}{\end{rblock}}
}

\newcommand{\article}[2]{
  \documentclass[#1pt,twoside,fleqn]{article}\usepackage{chicago}\std{#2} }
\newcommand{\nips}{
  \documentclass{article} \usepackage{nips01e,times} \stdpack\macros }
\newcommand{\ijcnn}{
  \documentclass[10pt,twocolumn]{/home/mt/usr/tex/ijcnn}
  \stdpack\macros
  \bibliographystyle{abbrv} 
}
\newcommand{\foga}{
  \documentclass{article} 
  \stdpack
  \usepackage{/home/mt/usr/tex/foga}
  \macros
}

\newcommand{\book}[2]{
  \documentclass[#1pt,twoside,fleqn]{book}\usepackage{chicago}\std{#2} }

\newcommand{\foils}[1]{
  \documentclass[12pt,fleqn]{article}
  \std{#1}
  \voffset -1cm  \addtolength{\textheight}{2cm}
  \renewcommand{\footskip}{2cm}
  \begin{document}
  \large
}
\newcommand{\landfoils}[1]{
  \documentclass[fleqn]{article}
  \stdpack
  \renewcommand{\baselinestretch}{#1}
  \renewcommand{\arraystretch}{1}
  \setlength{\hoffset}{-3.5cm}
  \setlength{\voffset}{-3.5cm}
  \setlength{\textwidth}{27cm}
  \setlength{\textheight}{19cm}
  \parindent 0ex
  \parskip 0ex 
  \pagestyle{empty}
  \macros
  \begin{document}
  \huge
}
\newcommand{\landfolien}[1]{
  \documentclass[fleqn]{article}
  \usepackage{german}
  \stdpack
  \renewcommand{\baselinestretch}{#1}
  \renewcommand{\arraystretch}{1.5}
  \setlength{\hoffset}{-5cm}
  \setlength{\voffset}{-1.5cm}
  \setlength{\textwidth}{27cm}
  \setlength{\textheight}{19cm}
  \parindent 0ex
  \parskip 0ex 
  \pagestyle{plain}
  \begin{document}
  \huge
}

\newcommand{\addressCologne}{
  Institute for Theoretical Physics\\
  University of Cologne\\
  50923 K\"oln---Germany\\
  {\tt mt@thp.uni-koeln.de}\\
  {\tt www.thp.uni-koeln.de/\~{}mt/}
}

\newcommand{\homepage}{{\tt www.neuroinformatik.ruhr-uni-bochum.de/PEOPLE/mt/}}
\newcommand{\email}{{\rm mt@neuroinformatik.ruhr-uni-bochum.de}}

\newcommand{\address}{\small\it
  Institut f\"ur Neuroinformatik,
  Ruhr-Universit\"at Bochum, ND 04,
  44780 Bochum---Germany\\
  \email
}

\newcommand{\mytitle}{
  \thispagestyle{empty}
  \rhead{\it Marc Toussaint---\today}
  \hrule height2pt
  \begin{list}{}{\leftmargin2ex \rightmargin2ex \topsep2ex }\item[]
    {\Large\bf \thetitle}
  \end{list}
  \begin{list}{}{\leftmargin7ex \rightmargin7ex \topsep0ex }\item[]
    Marc Toussaint \quad\today

    \address
  \end{list}
  \vspace{2ex}
  \hrule height1pt
  \vspace{5ex}
}

\newcommand{\contents}{{\small \parskip 0ex \tableofcontents \parskip 2ex }}

\newcommand{\sepline}{
  \begin{center} \begin{picture}(200,0)
    \line(1,0){200}
  \end{picture}\end{center}
}

\newcommand{\partsection}[1]{
  \vspace{5ex}
  \centerline{\sc\LARGE #1}
  \addtocontents{toc}{\contentsline{section}{{\sc #1}}{}}
}

\newcommand{\intro}[1]{\textbf{#1}\index{#1}}

\newtheorem{definition}{Definition}
\newtheorem{statement}{Statement}
\newtheorem{theorem}{Theorem}
\newtheorem{hypothesis}{Hypothesis}
\newenvironment{remark}{\noindent\emph{Remark.}}{}
\newenvironment{example}[1][]{\begin{block}[Example {#1}]}{\end{block}~}

\newcounter{parac}
\newcommand{\para}{\refstepcounter{parac}{\bf [{\roman{parac}}]}~~}
\newcommand{\Pref}[1]{[\emph{\ref{#1}}\,]}

\newenvironment{block}[1][]{{\noindent\bf #1}
\begin{list}{}{\leftmargin\blockindent \topsep-\parskip}
\item[]
}{
\end{list}
}

\newenvironment{rblock}{
\begin{list}{}{\leftmargin\blockindent \rightmargin\blockindent \topsep-\parskip}
\item[]
}{
\end{list}
}

\newenvironment{keywords}{\paragraph{Keywords}\begin{rblock}\small}{\end{rblock}}

\newenvironment{colpage}{
\addtolength{\columnwidth}{-3ex}
\begin{minipage}{\columnwidth}
\vspace{.5ex}
}{
\vspace{.5ex}
\end{minipage}
}

\newenvironment{enum}{
\begin{list}{}{\leftmargin3ex \topsep0ex \itemsep0ex}
\item[\labelenumi]
}{
\end{list}
}

\newenvironment{cramp}{
\begin{quote} \begin{picture}(0,0)
        \put(-5,0){\line(1,0){20}}
        \put(-5,0){\line(0,-1){20}}
\end{picture}
}{
\begin{picture}(0,0)
        \put(-5,5){\line(1,0){20}}
        \put(-5,5){\line(0,1){20}}
\end{picture} \end{quote}
}

\newcommand{\inputReduce}[1]{
{\sc\hspace{\fill} REDUCE file: #1}
}

\newcommand{\inputReduceInput}[1]{
  {\sc\hspace{\fill} REDUCE input - file: #1}
  \input{#1.tex}
}

\newcommand{\inputReduceOutput}[1]{
  {\sc\hspace{\fill} REDUCE output - file: #1}
}

\newcommand{\todo}[1]{{\bf[#1]}}

\newcommand{\macros}{
  \newcommand{\0}{{\hat 0}}
  \newcommand{\1}{{\hat 1}}
  \newcommand{\2}{{\hat 2}}
  \newcommand{\3}{{\hat 3}}
  \newcommand{\5}{{\hat 5}}

  \renewcommand{\a}{\alpha}
  \renewcommand{\b}{\beta}
  \renewcommand{\c}{\gamma}
  \renewcommand{\d}{\delta}
    \newcommand{\D}{\Delta}
    \newcommand{\e}{\epsilon}
    \newcommand{\g}{\gamma}
    \newcommand{\G}{\Gamma}
  \renewcommand{\l}{\lambda}
  \renewcommand{\L}{\Lambda}
    \newcommand{\m}{\mu}
    \newcommand{\n}{\nu}
    \newcommand{\N}{\nabla}
  \renewcommand{\k}{\kappa}
  \renewcommand{\O}{\Omega}
    \newcommand{\p}{\phi}
    \newcommand{\ph}{\varphi}
  \renewcommand{\P}{\Phi}
  \renewcommand{\r}{\varrho}
    \newcommand{\s}{\sigma}
    \newcommand{\Si}{\Sigma}
  \renewcommand{\t}{\theta}
    \newcommand{\T}{\Theta}
  \renewcommand{\v}{\vartheta}
    \newcommand{\X}{\Xi}
    \newcommand{\Y}{\Upsilon}

  \renewcommand{\AA}{{\cal A}}
    \newcommand{\GG}{{\cal G}}
  \renewcommand{\SS}{{\cal S}}
    \newcommand{\TT}{{\cal T}}
    \newcommand{\EE}{{\cal E}}
    \newcommand{\FF}{{\cal F}}
    \newcommand{\HH}{{\cal H}}
    \newcommand{\II}{{\cal I}}
    \newcommand{\KK}{{\cal K}}
    \newcommand{\LL}{{\cal L}}
    \newcommand{\MM}{{\cal M}}
    \newcommand{\NN}{{\cal N}}
    \newcommand{\CC}{{\cal C}}
    \newcommand{\PP}{{\cal P}}
    \newcommand{\QQ}{{\cal Q}}
    \newcommand{\RR}{{\cal R}}
    \newcommand{\UU}{{\cal U}}
    \newcommand{\YY}{{\cal Y}}
    \newcommand{\SOSO}{{\cal SO}}
    \newcommand{\GLGL}{{\cal GL}}

  \newcommand{\NNN}{{\mathbb{N}}}
  \newcommand{\ZZZ}{{\mathbb{Z}}}
  \newcommand{\RRR}{{\mathbb{R}}}
  \newcommand{\CCC}{{\mathbb{C}}}
  \newcommand{\one}{{{\bf 1}}}

  \newcommand{\<}{{\ensuremath\langle}}
  \renewcommand{\>}{{\ensuremath\rangle}}
  \newcommand{\Aut}{{\rm Aut}}
  \newcommand{\cor}{{\rm cor}}
  \newcommand{\corr}{{\rm corr}}
  \newcommand{\cov}{{\rm cov}}
  \newcommand{\sd}{{\rm sd}}
  \newcommand{\tr}{{\rm tr}}
  \newcommand{\lag}{\mathcal{L}}
  \newcommand{\inn}{\rfloor}
  \newcommand{\lie}{\pounds}
  \newcommand{\longto}{\longrightarrow}
  \newcommand{\speer}{\parbox{0.4ex}{\raisebox{0.8ex}{$\nearrow$}}}
  \renewcommand{\dag}{ {}^\dagger }
  \newcommand{\h}{{}^\star}
  \newcommand{\w}{\wedge}
  \newcommand{\too}{\longrightarrow}
  \newcommand{\To}{\Rightarrow}
  \newcommand{\Too}{\;\Longrightarrow\;}
  \newcommand{\ow}{\stackrel{\circ}\wedge}
  \newcommand{\feed}{\nonumber \\}
  \newcommand{\comma}{\; , \quad}
  \newcommand{\period}{\; . \quad}
  \newcommand{\del}{\partial}
  \newcommand{\point}{$\bullet~~$}
  \newcommand{\doubletilde}{
  ~ \raisebox{0.3ex}{$\widetilde {}$} \raisebox{0.6ex}{$\widetilde {}$} \!\!
  }
  \newcommand{\topcirc}{\parbox{0ex}{~\raisebox{2.5ex}{${}^\circ$}}}
  \newcommand{\sym}{\topcirc}

  \newcommand{\half}{\frac{1}{2}}
  \newcommand{\third}{\frac{1}{3}}
  \newcommand{\fourth}{\frac{1}{4}}

  \renewcommand{\_}{\underset}
  \renewcommand{\^}{\overset}

  \renewcommand{\small}{\footnotesize}
}

\newcommand{\argmax}[1]{\text{arg}\underset{#1}\max}
\newcommand{\argmin}[1]{\text{arg}\underset{#1}\min}
\newcommand{\kld}[2]{D\!\left(\,#1\,|\!|\,#2\,\right)}

\newcommand{\pathmt}{./}
\newcommand{\basepath}{./}
\newcommand{\setpath}[1]{\renewcommand{\pathmt}{#1}\renewcommand{\basepath}{#1}}
\newcommand{\pathinput}[2]{
  \renewcommand{\pathmt}{\basepath #1}
  \input{\pathmt #2} \renewcommand{\pathmt}{\basepath}}

\newcommand{\hide}[1]{[\small #1 \normalsize]}
\newcommand{\color}[2][1]{}

\article{10}{1}

\newcommand{\df}{\d\!f}

\title{\Large\textbf{On model selection and the disability of neural networks\\ to decompose tasks}}

\author{\normalsize Marc Toussaint\\
 \sizeix Institut f\"ur Neuroinformatik, Ruhr-Universit\"at Bochum\\
\sizeix 44780 Bochum, Germany\\
\textit{\sizeix Marc.Toussaint@neuroinformatik.ruhr-uni-bochum.de}}

\date{}

\begin{document}


\twocolumn[\mytitle]\thispagestyle{fancy}

\rhead{\it Proceedings of the International Joint Conference on Neural
  Networks (IJCNN 2002)}

\begin{abstract}%
  A neural network with fixed topology can be regarded as a
  parametrization of functions, which decides on the correlations
  between functional variations when parameters are adapted. We
  propose an analysis, based on a differential geometry point of view,
  that allows to calculate these correlations. In practise, this
  describes how one response is unlearned while another is trained.
  Concerning conventional feed-forward neural networks we find that
  they generically introduce strong correlations, are predisposed to
  forgetting, and inappropriate for task decomposition.  Perspectives
  to solve these problems are discussed.



\end{abstract}

\section{Introduction}\label{intro}

Following Kerns et al. \citeyear{kearns:95}, the problem of model
selection may be defined as follows: Given a finite set of data
points, find a function (or conditional probability distribution, also
called hypothesis) such that the expected generalization error is
minimized.  Typically, the search space $\FF$ (the space of functions
or conditional probability distributions) is assumed to be organized
as a nested sequence of subspaces $\FF_1 \subseteq ..  \subseteq \FF_d
\subseteq .. \subseteq \FF$ of increasing complexity.  For instance,
the index $d$ may denote the number of parameters or the
Vapnik-Chervonenkis dimension \cite{vapnik:95}. Finding the function
with minimal generalization error then amounts to finding the
appropriate sub-search-space before applying ordinary optimization
schemes.  Many approaches introduce a penalty term related to
complexity which has to be minimized together with the training error.
Penalty terms are, for example, the number of parameters of the model,
the number of \emph{effective} model parameters, the
Vapnik-Chervonenkis dimension, or the description length
\cite{akaike:74,amari:93,moody:91,rissanen:78,vapnik:95}. An
alternative based on geometric arguments is presented by
Schuurmans \citeyear{schuurmans:97}.

The emphasis of our investigations is different to these classical
approaches.  The choice of a specific model (e.g., a neural network)
to represent a function has \emph{two} implications: it defines the
space $\FF_d$ of representable functions, but it also defines a
\emph{parametrization} of this space, where parametrization is not
meant in the sense of `finding parameters' but in the sense of
introducing coordinates on that space, i.e., introducing a mapping
$\P:\, \RRR^m \to \FF_d$ from some coordinate space $\RRR^m$ onto the
sub-search-space. To omit confusion, we use the term \emph{model
  class} for the sub-search-space $\FF_d$, and \emph{model
  parametrization} for the parametrization $\P$ of this
sub-search-space.  For example, an artificial neural network with $m$
free parameters, fixed topology, and fixed activation functions
defines a model class (the subspace of functions it can
realize---which, if the topology is appropriate, includes an
approximation of any function \cite{hornik:89}) but it also defines a
model parametrization (the mapping from its parameters to the
corresponding function).

Our emphasis is on the implications of a specific model
parametrization instead of the choice of a certain model class. It is
important to have a closer look at this parametrization in order to
allow for an analytical description of the adaptation dynamics, rather
than just analyzing the complexity of a model class. In particular,
the precise relation between variations of parameters and functional
variations of the system is of fundamental interest because it
decides, e.g., on the way of ``extrapolation'', or on how the system
forgets previously learned data. This relation can be derived from the
model parametrization and our goal is to extract such features
analytically.  We focus on forgetting as a specific character of
adaptation dynamics and develop an analysis of the model
parametrization that allows to approximate the rate of forgetting.
This analysis is based on a differential geometry point of view and is
related to a large pool of research, including the discussions of
\emph{cross-talk} \cite{jacobs:90} and \emph{catastrophic forgetting}
\cite{french:99}, the information geometry point of view on parameter
adaptation \cite{amari:00}, and perfectly analogous ideas in the
context of evolutionary adaptation \cite{toussaint:01}.  Section
\ref{ana} includes a discussion of these relations.

We apply our method of analyzing the model para\-metrization on the
class of standard feed-forward neural networks (FFNNs). We find that
the variety of FFNNs with arbitrary topology is actually not a great
variety with respect to certain characters of the model
parametrization. In particular, FFNNs gnerically introduce strong
correlations between functional variations and thereby are predisposed
to forget previously learned data.  Hence, using FFNNs as a function
model means a limitation---not with respect to representable functions
but with respect to learning characteristics. A simple example
compares a standard FFNN with a network that includes competitive
interactions.  The results validate our analytical predictions and
illustrate their implications. We conclude that a generalization of
the class of FFNNs is necessary and that the introduction of
competitive interactions between neurons is a promising approach to
solve these problems.

Section \ref{def} will introduce to the formalism our investigations
are based on and, in section \ref{ana}, we describe the analysis of
the model parametrization. Section \ref{emp} presents the examples
and in section 5 we give an outlook concerning the evolutionary
perspective on model selection and discuss the relevance of the
limitedness of FFNN models. The conclusion follows up.

\section{Definitions}\label{def}

\subsection{The functional point of view}

Let $\FF$ be the search space. Here, $\FF$ shall be the space of all
functions mapping from a finite space $X$ to $Y \subseteq \RRR^n$.
However, all results can be transferred to the search space of
conditional probabilities, as we discuss below.

The space of functions $f:\, X \to Y$ can be written as $Y^X$, which
is isomorphic to $\RRR^{n\cdot|X|}$. Thus, let a function $f \in
Y^X$ be represented by $n\!\cdot\!|X|$ components $f^a \in \RRR$,
where the index $a$ refers to a specific point in $X$ \emph{and} a
$Y$-dimension. (The components $f^a$ may be regarded as entries of a
lookup-table representation of $f$.) On this representation, we
describe an online adaptation step as a probabilistic transition to a
new function as follows: Assume that adaptation is initiated by the
observation of a target value $t^a$ for a functional component $f^a$.
A transition occurs as a variation $\df \in \RRR^{n|X|}$ with
probability $p(\df\, |\, f^a,t^a)$. The interesting point is that
functional components of which no target value has been observed may
vary as well. Let $a$ be a random variable and consider the density
$p(\df)=p(\df\, |\, f^a,t^a)\, p(a)$. We will refer to the respective
covariance between two variation components as the \emph{functional
  covariance matrix}
\begin{align}
  C^{bc} := \cov_{p(\df)}(\df^b,\df^c) \;.
\end{align}

This matrix is a first order description of how the adaptation of the
observed functional component results in a \emph{coadaptation} of a
functional component which has not been observed. For example,
assuming a linear dependence between $\df^a$ and $\df^b$, we have
$\df^b \stackrel{\cdot}= \<\df^b\> + \frac{C^{ab}}{\s^2}\, \big( \df^a
- \<\df^a\>\big)$, where $\s^2$ is the variance of $\df^a$. Whether
this coadaptation is desirable or not depends on the problem.
Coadaptation is also an explicit description of the ``way of
generalization''\footnote{By ``way of generalization'' we do not refer
  to the generalization error but to the way of extrapolation from
  observed data to unobserved.}: unobserved functional components
(i.e., the functional response on stimuli that have not been observed)
are coadapted depending on the adaptation of observed functional
components.  In general, one would like to choose from a variety of
different coadaptation schemes, i.e., one would like to select a model
from a variety of models with different kinds of coadaptation. We will
find that this refers to the selection of a model parametrization.

When the set of functional components can be separated in two disjoint
subsets such that $C^{ab}$ vanishes for two components $f^a$ and $f^b$
of different subsets, then we speak of \emph{adaptation
  decomposition}. During online learning, adaptation decomposition
means that the development of two such components during successive
adaptation is not correlated. In terms of homogeneous Markov
processes, successive adaptation is described by the transition
probability $p(\df\, |\, f^a,t^a)$ (assuming that the draw of $a$ from
$p(a)$ is independent at each time), and adaptation is decomposed if
$p(\df^a,\df^b)=p(\df^a)\, p(\df^b)$.

\subsection{The parameter point of view}

We now address the \emph{modeling} of functions. Let $\P$ be a
$m$-dimensional, differentiable parametrization of a subset $\P(W)$
of functions:
\begin{align}
& \P:\, W \to \FF \comma W\subseteq\RRR^m \;,\\
& \P(W) := \bigcup_{w \in W} \{\P(w)\} \quad \subseteq \FF \;.
\end{align}
We call $\P$ the \emph{model parametrization} and $\P(W)$ the
\emph{model class}.  In terms of differential geometry, $\P$ is the
inverse of a coordinate map (or chart, or atlas) for $\P(W)$. Since
this map is differentiable, it induces a metric on $\P(W)$ if one on
$W$ is given and vice versa. We define the \emph{functional metric}
$g^{ab}(w)$ on $\P(W)$ as the lift of the Euclidean metric on $W$,
\begin{align}
g^{ab}(w) := \sum_i \frac{d \P(w)^a}{d w^i}\, \frac{d \P(w)^b}{d w^i} \;;
\label{funcMet}
\end{align}
and we define the \emph{parameter metric} $g_{ij}(w)$ on $W$ (actually
on the \emph{dual} tangent spaces of $W$) as the pull-back of the
Euclidean metric on $\P(W)$,
\begin{align}
g_{ij}(w) := \sum_a \frac{d \P(w)^a}{d w^i}\, \frac{d \P(w)^a}{d w^j} \;.
\label{parMet}
\end{align}
As usual in differential geometry, the metrics depend on the locality
given by $w$. These metrics describe the relation between parameter
variations and functional variations as we explore in more detail in
the next section.

\section{Analysis of the model para\-metrization}\label{ana}

In the previous section we defined the correlation matrix $C^{ab}$ on
the functional level. Now we analyze what the choice of a model
parametrization $\P$ implies on this functional level. Given $\P$ and
parameters $w$, we write $f^a=\P(w)^a$. Assume that a target $t^a$ was
observed and adaptation of the parameters takes place by a gradient
descent,
\begin{align}
\d w^i = 2 \a\; \frac{d f^a}{d w^i}\, (t^a - f^a) \;,
\end{align}
which corresponds to the gradient of the squared error multiplied by
an adaptation rate $\a$. In first order approximation, this induces a
functional variation
\begin{align}
\df^b 
  = 2 \a\; \sum_i \frac{d f^b}{d w^i}\, \d w^i 
  = 2 \a\; g^{ab}\, (t^a - f^a) \;,
\label{deltaF}
\end{align}
using definition (\ref{funcMet}). Thus, the functional metric $g^{ab}$
describes the variation of a functional component $f^b$ when $t^a$ is
observed.  This gives a first order description of coadaptation and of
how the model generalizes the experience of a target value $t^a$ in
order to adapt also functional components $f^b$. In this approximation
the functional covariance reads
\begin{align}\label{covari}
C^{bc} = 4 \a^2\; \sum_a p(a)\; g^{ba}\, g^{ca}\, (t^a - f^a)^2 -\<\df^a\>\<\df^b\>\;.
\end{align}
To discuss this expression, let us assume that the second term
vanishes, $\<\df^a\>\<\df^b\>=0$. Concerning the first term, the
product $g^{ba}\, g^{ca}$ vanishes for all $a$ if and only if the
functional metric is a block matrix and $b$ and $c$ refer to different
blocks:
\begin{align*}
g^{ab} = \left(\begin{array}{cc}
           A \in \RRR^{\m\times\m} & 0\\
           0 & B\in \RRR^{\n\times\n}
    \end{array}\right) \comma b\le\m \comma c>\m \;,
\end{align*}
where $A$ and $B$ are arbitrary symmetric matrices and
$\m+\n=n\cdot|X|$. Thus, adaptation is decomposed into two subsets of
functional components exactly if the functional metric is a block
matrix and the functional component subsets correspond to these
blocks.\footnote{Note the relation to group theory: A group
  representation is said to be reducible if all group generators can
  be represented as a block matrix (such that all of them fit in the
  same block template). On this basis, physics defines the notion of
  an elementary particle as corresponding to an irreducible
  representation, whereas physical systems that correspond to a
  reducible representation (a block matrix) are considered as
  \emph{composed} of particles. A system of which the adaptation
  dynamics (instead of physical interactions) can be decomposed in the
  sense of a block matrix can analogously be thought of as composed of
  subsystems.

  More formally, the observation of a target $t^a$ can be identified
  with an element of a group that applies on the functional
  components. Adaptation dynamics is now interpreted as successive
  application of group elements. The group representation (i.e., the
  way the group elements apply on the functional components) is
  determined by the model parametrization. If adaptation is
  decomposed, this representation is reducible.}

\subsection{Reference to related research}

\paragraph{Cross-talk.}
The inspiring work by Jacobs et al. \citeyear{jacobs:90} discusses the
implication of the choice of a multi-expert model on the learning
speed and generalization behavior. They formulate the idea of spatial
and temporal crosstalk, which denotes the statistical dependence
between the states of two different neurons or between the states of a
neuron at two different times. In our formalism, this crosstalk is
captured by the functional covariance---spatial for two indices $a$
and $b$ belonging to the same input $x \in X$, and temporal for two
indices of different input. They argue that such a crosstalk may be
undesirable and is avoided by explicitly separating neurons in
disjoint experts. As we will see below, selecting a multi-expert model
is a very intuitive way to explicitly declare an independence of
functional components and realize decomposed adaptation. In fact, the
separation into experts corresponds to a block matrix type functional
metric. (If the gating is also adaptive, the functional metric is
actually not a completely clean block matrix.)

In the context of artificial neural networks, the term
\emph{catastrophic forgetting} has been used to describe negative
effects of coadaptation. See \cite{french:99} for a review.

\paragraph{Information geometry.}
The methods applied in this paper are related to information geometry.
Let $Y=S_\n=[0,1]^{2^\n-1}$ be the $2^\n-1$ dimensional manifold of
probability distributions over $\{0,1\}^\n$ as defined by Arami
\citeyear{amari:00}. Then, the search space $\FF$ of mappings $X \to Y$
is the space of all conditional probabilities $p(y|x)$, $x\in X, y\in
Y$. Usually, one assumes the Fisher metric on $\FF$, not the
Euclidean. Thus, we would have to change the definition (\ref{parMet})
of the parameter metric into
\begin{align}
g_{ij}(w)=E\left[
  \frac{\del\log p(x,y;w)}{\del w^i}\;
  \frac{\del\log p(x,y;w)}{\del w^j}\right]\;,
\end{align}
where $E[.]$ denotes the expectation and $p(x,y;w)=p(y|x;w)\,p(x)$,
$p(y|x;w)=\P(w) \in \FF$. Arami \citeyear{amari:98} uses this metric
to define the natural gradient descent on the parameter space (which
actually is the covariant derivative instead of the contravariant).
The use of the natural gradient can also be motivated by a
spatio-temporal decorrelation \cite{choi:00}.

\paragraph{Evolutionary computation.}
It seems that in the field of evolutionary computation the discussion
of the covariance structure in the search space is much more
elaborated than in the field of neural computation (see
\citeNP{toussaint:01}). Roughly speaking, the goal of evolutionary
computation is to maximize the probability of good mutations during
evolutionary search. Eventually, fitness requires some phenotypic
traits to be mutated in correlation. Such correlations (coadaptation)
may be modeled explicitly in the search density of evolutionary
algorithms \cite{baluja:97,hansen:01,muehlenbein:99,pelikan:99}.
Alternatively, they may be induced implicitly by the choice of a good
parametrization of phenotypic traits---by a genotype-phenotype
mapping, which is in perfect analogy to the model parametrization
$\P$. Many research efforts focus on the choice or the understanding
of the genotype-phenotype mapping
\cite{stephens:99,toussaint:01,wagner:96}.  In this view, functional
components $f^a$ may be compared to phenotypic traits, whereas
parameters relate to the genotype.

\section{Example}\label{emp}

Our test of the learning behavior is very simple: a regression of only
two patterns in $\{0,1\}^3$ has to be learned by mapping the first
pattern on $+1$ and the second on $-1$. However, we impose that these
patterns have to be learned \emph{online} where they alternate only
after they have been exposed for $100$ times in
succession.\footnote{This task is not meant as a performance test but
  as an experimental setup to test our analytical methods. However,
  similar effects of learning and unlearning occur in online learning
  when a specific response is unlearned during the course of training
  other responses for several time steps. In real world simulations it
  is also plausible that stimuli remain unchanged for many time
  steps.} We test two systems on this task: a standard feed-forward
neural network as described in detail in table \ref{FFN}, and a system
that involves a softmax layer as described in table \ref{monet}. The
parameters of both systems are initialized randomly by the normal
distribution $\NN(0,0.1)$ around zero with standard deviation $0.1$.
The two patterns were chosen as $110$ and $010$. Learning is realized
by a slow gradient descent with adaptation rate $2\cdot10^{-3}$ and
momentum $0.5$. The metric components are calculated from the
gradients.

\begin{table}[t]
  \fbox{\begin{colpage} The feed-forward neural network we investigate
    here is 3-4-1-layered; layers are completely connected; the output
    neurons are linear, the hidden ones implement the sigmoid
    $\frac{1}{1+\exp(-10\, x)}$; only the hidden neurons have bias
    terms.
\end{colpage}}
\caption{The Standard model}
\label{FFN}
\end{table}

\begin{table}[t]
  \fbox{\begin{colpage} The softmax model is the same as the standard
    model with the exception that the four neurons in the hidden layer
    \emph{compete} for activation: their output activations $y_i$ are
    given by
\begin{align}
y_i=\frac{e^{30\, x_i}}{X} \comma
&x_i=\sum_{j \in\, \text{input}} w_{ij} y_j + w_i \;,\feed
&X=\sum_{i \in\, \text{hidden}} e^{30\, x_i} \;.
\label{gating}
\end{align}
Here, $w_{ij}$ and $w_i$ denote weight and bias parameters. The
exponent factor $30$ may be interpreted as rather low temperature,
i.e., high competition. The calculation of the gradient is a little
more involved than ordinary back-propagation but straightforward and
of same computational cost (see \cite{toussaint:02b}).
\end{colpage}}
\caption{The Softmax model}
\label{monet}
\end{table}

\begin{figure}[t]\center
\psfrag{g_00}{\!\small $g^{00}$}
\psfrag{g_01}{\!\small $g^{01}$}
\psfrag{g_11}{\!\small $g^{11}$}
\psfrag{empirical}{\small measured}
\psfrag{estimated}{\small calculated}
\psfrag{trained}{\small trained}
\psfrag{untrained}{\small untrained}
\includegraphics[width=\columnwidth]{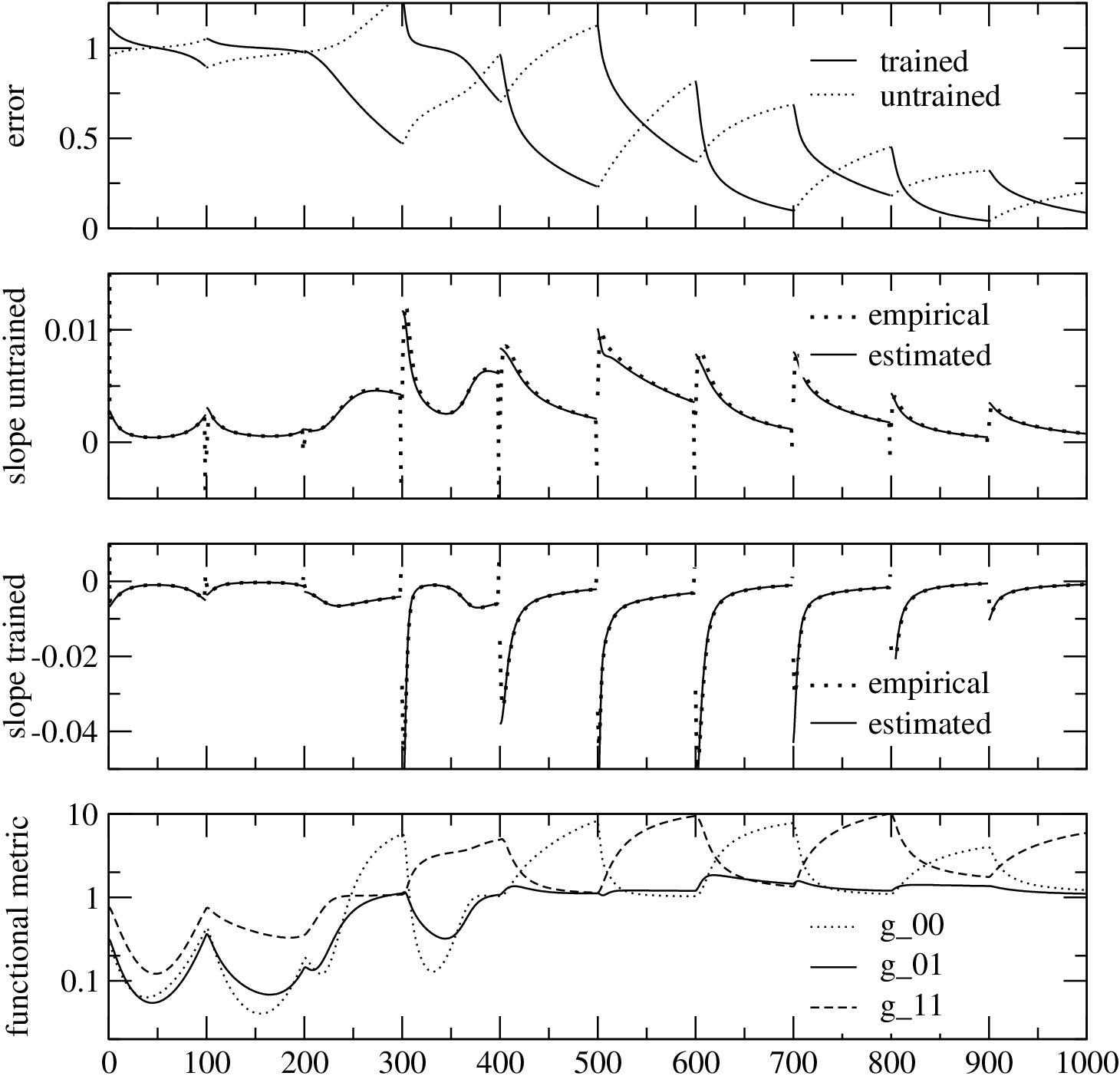}
\caption{\emph{Test of the standard model.}\newline
  \small For all four graphs the abscissa denotes the time
  step.\newline \emph{Top:} The learning curves (errors) with respect
  to both patterns are displayed. Only one of the patterns is
  trained---alternating every 100 time steps.  The error of the
  untrained patterns increases.\newline \emph{Second:} The slope
  (change of error per time step) of the untrained learning curve is
  displayed. The dotted line refer to the measured slope of the upper
  curve, the normal line is calculated according to equation
  (\ref{deltaF}).\newline \emph{Third:} The slope (measured and
  calculated) of the trained learning curve.\newline \emph{Bottom:}
  The three components of the functional metric $g^{00}$, $g^{01}$,
  $g^{11}$ are displayed in logarithmic scale. In particular the
  cross-component $g^{01}$ is clearly non-vanishing.}
\label{curvesStd}
\end{figure}

\begin{figure}[t]\center
\psfrag{g_00}{\!\small $g^{00}$}
\psfrag{g_01}{\!\small $g^{01}$}
\psfrag{g_11}{\!\small $g^{11}$}
\psfrag{empirical}{\small measured}
\psfrag{estimated}{\small calculated}
\psfrag{trained}{\small trained}
\psfrag{untrained}{\small untrained}
\includegraphics[width=\columnwidth]{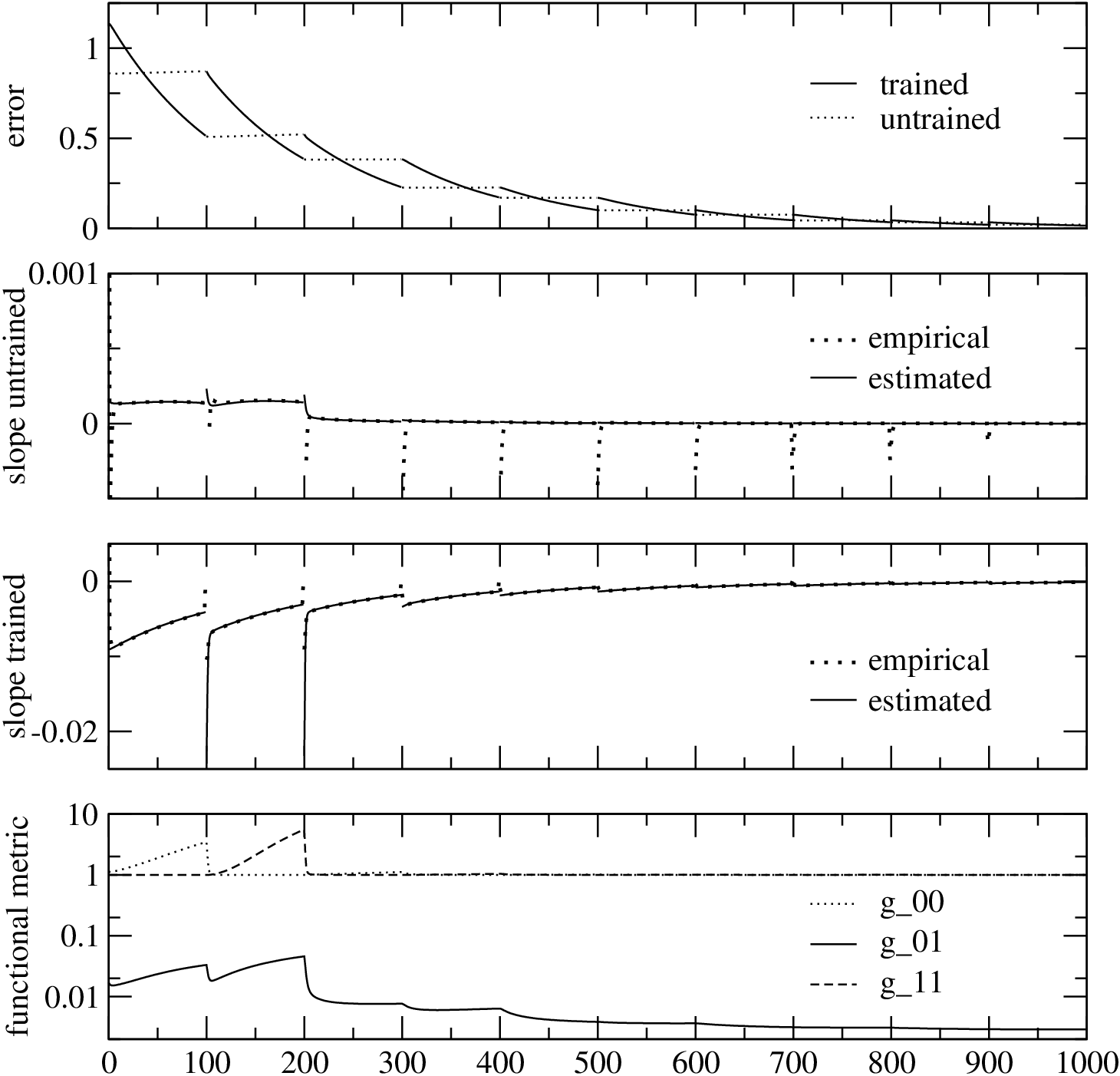}
\caption{\emph{Test of the softmax model.} \newline
  \small\emph{Top:} The learning curves (errors) with respect to both
  patterns are displayed. The untrained patterns is scarcely
  forgotten.\newline \emph{Second:} The slope (measured and
  calculated) of the untrained learning curve nearly vanishes.\newline
  \emph{Third:} The slope (measured and calculated) of the trained
  learning curve.\newline \emph{Bottom:} The three components of the
  functional metric $g^{00}$, $g^{01}$, $g^{11}$ (in logarithmic
  scale). The cross-component $g^{01}$ is small, it decreases
  significantly at time step 200.}
\label{curvesSoft}
\end{figure}

Please see Figures \ref{curvesStd} and \ref{curvesSoft} for the
results. For the standard neural model we observe some forgetting of
the untrained pattern during the training of the other. For the
softmax model, the error of the untrained pattern hardly increases.
The rate of forgetting, given by the slope of the error curve, is well
described by equation (\ref{deltaF}) and demonstrated by the graphs in
the middle. The bottom graphs display the functional metric components
and generally exhibit that the cross-component $g^{01}$, which is
responsible for coadaptation and forgetting, is quite large for the
standard model compared to the softmax model. Further, the softmax
model seems to learn the adaptation decomposition, as defined in
section \ref{def}, after the 200th time step. All these results reveal
that the standard model is not well-suited to solve the simple task
given and that the analysis of the model's functional metric provides
a formal way of understanding this phenomenon. Remarkably also, the
components $g^{00}$ and $g^{11}$ become significantly greater than $1$
during the training phase of the respective functional component. By
equation (\ref{deltaF}), this means that the ``effective'' adaptation
rate is larger than $2\cdot10^{-3}$.

\begin{figure}[t]\center
\psfrag{g00}{\!\small $g^{00}$}
\psfrag{g01}{\!\small $g^{01}$}
\psfrag{g11}{\!\small $g^{11}$}
\includegraphics[width=\columnwidth]{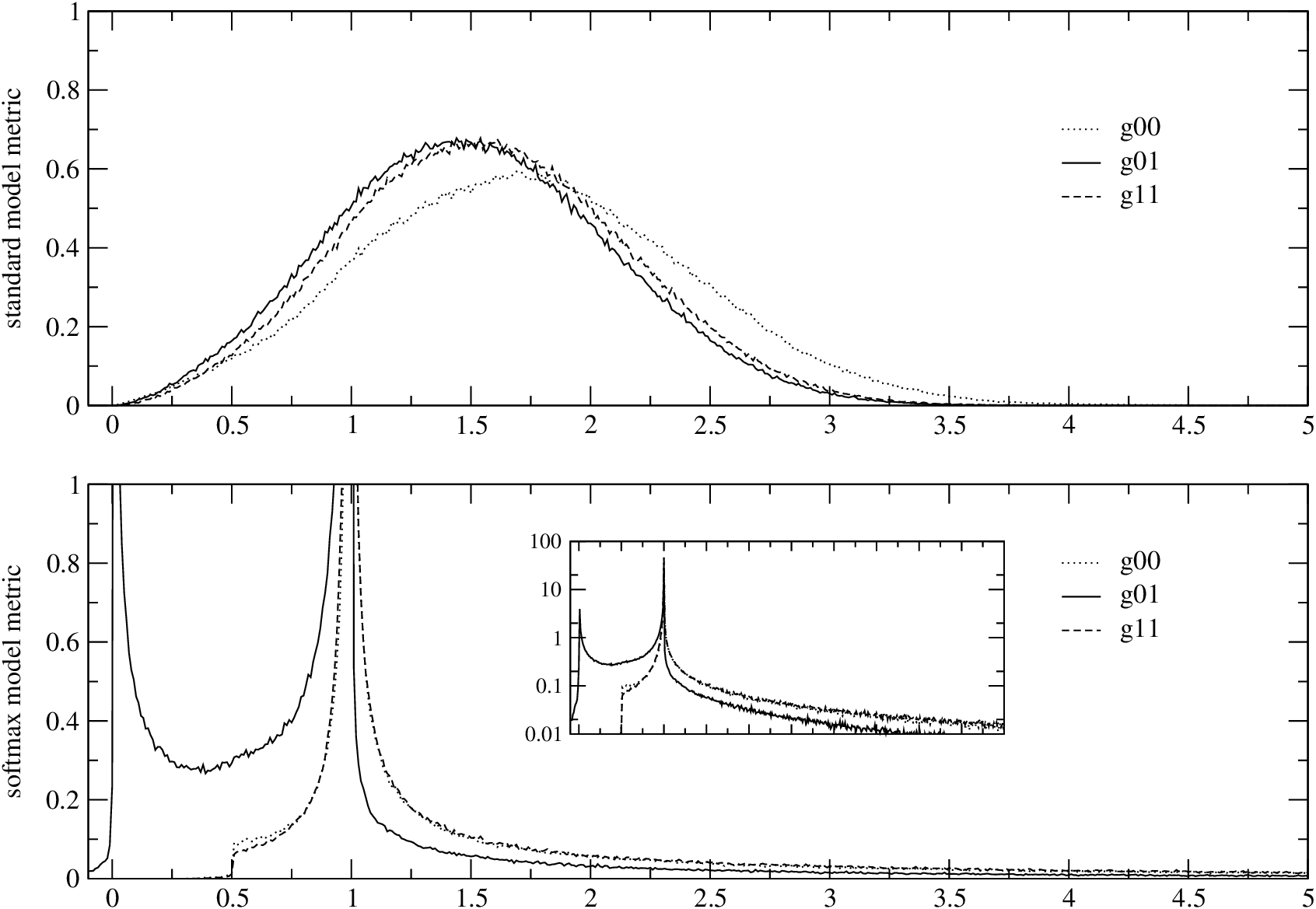}
\caption{\emph{Distribution of metric components.}\newline
  The distribution was calculated as a histogram of $1$ million
  samples by using bins of size $\frac{1}{100}$. The ordinate is
  scaled in ``percent of samples that fell into the bin''.\newline
  \emph{Top:} The standard model. The probability of vanishing
  cross-component $g^{01}$ is vary small.\newline \emph{Bottom:} The
  softmax model. The inset graph is in logarithmic scale. The
  probability of vanishing cross-component $g^{01}$ is fairly high.}
\label{distri}
\end{figure}

One might object that the results given above rely on the random
initialization and on the specific task we chose. To analyze both
types of models in a more general way we perform another test. We
investigate the distribution of the functional metric components when
parameters are normally distributed by $\NN(0,0.1)$.  Figure
\ref{distri} shows the distributions for both models. Clearly, the
standard model exhibits a Gauss-like distribution of the
cross-component $g^{01}$ with mean around $1.5$; a vanishing
cross-component $g^{01}$ is not very likely.  On the other hand, the
softmax model exhibits two strong peaks at $g^{01}=0$ and $g^{01}=1$,
such that the probability for $g^{01}<0.1$ is larger than 10\%. These
distributions are generic properties of the two models.

\section{Toward evolutionary model selection}

Finally, the question of how to select an appropriate model has not
yet been addressed. As discussed in the introduction, classical
approaches to model selection commonly introduce a penalty term in
order to reduce the model's complexity. Following this tradition we
could introduce a penalty term that reduces forgetting. Consider
\begin{align}
\sum_{ab} (g^{ab})^2 - \sum_a (g^{aa})^2 \;.
\end{align}
This is a measure of the cross-components in the functional metric.
Unfortunately, we cannot present any experiments with this model
selection criterion here. This approach is postponed to future
research.

The original motivation for this work, though, was not to develop a
new model selection criterion as given by the above penalty term.
Instead we believe that the evolution of neural networks, as it
recently became an elaborated branch of research (see \cite{yao:99}
for a review), is actually a promising method of model selection.
However, most of these approaches focus on standard neural models,
i.e., the evolutionary search space is the space of ordinary
feed-forward neural networks (FFNNs) with arbitrary topology. The
belief is that the variety of topologies offers a variety of
functionally different models. The present paper is a critique of this
belief because it supports that the functional metric inherent of
FFNNs comprises significantly non-vanishing cross-components. This
implies that the variety of FFNNs with arbitrary topology is actually
not a great variety with respect to the functional metric. E.g., it
hardly includes models with vanishing cross-components and low rate of
forgetting. In conclusion, the search space has to be generalized to
contain also models with arbitrary functional metric in order to allow
for the selection of more optimal models. The presented softmax model
involving competitive interactions between neurons is a step in this
direction, but much motivation is left for future research toward the
generalization of the model search space and evolutionary methods to
select good models from this great variety. The model presented in
\cite{toussaint:02b} is one approach.

\section{Conclusion}

We developed a new analytical approach to characterize a function
model and describe its learning properties. We focussed on functional
correlations in the adaptation process and derived the relation to the
functional metric of the model parametrization. The analysis can in
principal be applied on any kind of differentiable model (also
probabilistic, when formulated in terms of information geometry). Our
empirical studies illustrate the approach and demonstrate that
conventional neural network models are rather limited with respect to
their adaptation behavior: a task separation, i.e., decorrelated
adaptation to decorrelated data, is hardly possible. In contrast, a
model involving competitive interactions is more predisposed for task
decomposition. Thus, as we pointed out in the previous section, the
evolutionary approach to model selection should generalize the search
space to include not only standard feed-forward neural networks, but
also models with arbitrary functional metrics, e.g., by allowing for
competitive interactions.

\subsection*{Acknowledgment}

The author acknowledges support by the German Research Foundation DFG
under grant \emph{SoleSys} SE 251/41-1.

\small
\bibliography{/home/mt/bibtex/bibs}
\end{document}